\documentclass[12pt]{article}
\usepackage{epsfig}

\newcommand\pubnumber{CERN-TH/2001-015}
\newcommand\pubdate{January 22, 2001}
\newcommand\hepnumber{hep-ph/0101248}
\def\csumb{TH Division, CERN, CH-1211 Geneva 23, Switzerland}
\def\support{} 

\def\Title#1{\begin{center} {\Large\bf #1 } \end{center}}
\def\Author#1{\begin{center}{ \sc #1} \end{center}}
\def\Address#1{\begin{center}{ \it #1} \end{center}}

\newcommand\pubblock{\rightline{\begin{tabular}{l} \pubnumber\\
         \pubdate\\ \hepnumber \end{tabular}}}
\newenvironment{Abstract}{\begin{quotation}  }{\end{quotation}}
\newenvironment{Presented}{\begin{quotation} \begin{center} 
             Presented at the\end{center}
      \begin{center}\begin{large}}{\end{large}\end{center} \end{quotation}}
\def\Acknowledgments{\bigskip  \bigskip \begin{center}
          \large\bf Acknowledgments\end{center}}

\makeatletter
\def\section{\@startsection{section}{0}{\z@}{5.5ex plus .5ex minus
 1.5ex}{2.3ex plus .2ex}{\large\bf}}
\def\subsection{\@startsection{subsection}{1}{\z@}{3.5ex plus .5ex minus
 1.5ex}{1.3ex plus .2ex}{\normalsize\bf}}
\def\subsubsection{\@startsection{subsubsection}{2}{\z@}{-3.5ex plus
-1ex minus  -.2ex}{2.3ex plus .2ex}{\normalsize\sl}}

\renewcommand{\@makecaption}[2]{%
   \vskip 10pt
   \setbox\@tempboxa\hbox{\small #1: #2}
   \ifdim \wd\@tempboxa >\hsize     
       \small #1: #2\par          
     \else                        
       \hbox to\hsize{\hfil\box\@tempboxa\hfil}
   \fi}

 \def\citenum#1{{\def\@cite##1##2{##1}\cite{#1}}}
 
\newcount\@tempcntc
\def\@citex[#1]#2{\if@filesw\immediate\write\@auxout{\string\citation{#2}}\fi
  \@tempcnta\z@\@tempcntb\m@ne\def\@citea{}\@cite{\@for\@citeb:=#2\do
    {\@ifundefined
       {b@\@citeb}{\@citeo\@tempcntb\m@ne\@citea\def\@citea{,}{\bf ?}\@warning
       {Citation `\@citeb' on page \thepage \space undefined}}%
    {\setbox\z@\hbox{\global\@tempcntc0\csname b@\@citeb\endcsname\relax}%
     \ifnum\@tempcntc=\z@ \@citeo\@tempcntb\m@ne
       \@citea\def\@citea{,}\hbox{\csname b@\@citeb\endcsname}%
     \else
      \advance\@tempcntb\@ne
      \ifnum\@tempcntb=\@tempcntc
      \else\advance\@tempcntb\m@ne\@citeo
      \@tempcnta\@tempcntc\@tempcntb\@tempcntc\fi\fi}}\@citeo}{#1}}
\def\@citeo{\ifnum\@tempcnta>\@tempcntb\else\@citea\def\@citea{,}%
  \ifnum\@tempcnta=\@tempcntb\the\@tempcnta\else
  {\advance\@tempcnta\@ne\ifnum\@tempcnta=\@tempcntb \else\def\@citea{--}\fi
    \advance\@tempcnta\m@ne\the\@tempcnta\@citea\the\@tempcntb}\fi\fi}
\makeatother

\def\lsim{\mathrel{\raise.3ex\hbox{$<$\kern-.75em\lower1ex\hbox{$\sim$}}}}
\def\gsim{\mathrel{\raise.3ex\hbox{$>$\kern-.75em\lower1ex\hbox{$\sim$}}}}

\begin{document}
\begin{titlepage}
\pubblock

\vfill
\def\thefootnote{\fnsymbol{footnote}}
\Title{Conformal Expansions:  \\[5pt] A Template for QCD Predictions}
\vfill
\Author{Johan Rathsman\support}
\Address{\csumb}
\vfill
\begin{Abstract}
The use of conformal expansions for predictions in quantum chromodynamics 
is discussed as a way to avoid renormalization scheme and scale ambiguities, 
as well as factorial growth of perturbative coefficients due to renormalons.
Special emphasis is given to the properties of an
assumed skeleton expansion and its relation to the Banks-Zaks expansion.
The relation of BLM scale-setting to the skeleton expansion is also
discussed and new criteria for the applicability of BLM scale-setting 
are presented.

\end{Abstract}
\vfill
\begin{Presented}
5th International Symposium on Radiative Corrections \\ 
(RADCOR--2000) \\[4pt]
Carmel CA, USA, 11--15 September, 2000
\end{Presented}
\vfill
\end{titlepage}
\def\thefootnote{\arabic{footnote}}
\setcounter{footnote}{0}


\section{Introduction}

The apparent freedom in choosing the renormalisation scale and scheme
for perturbative calculations of observables in quantum chromodynamics 
(QCD) introduces theoretical uncertainties which, if taken literally,
prohibit absolute predictions beyond a qualitative level.
The renormalisation scheme dependence can be solved by using a fixed 
reference scheme or, equivalently, by relating measurements of different 
observables to each other. The $\overline{\mbox{MS}}$ scheme 
is the simplest choice from a calculational point of view but
the question then arises if there exists a preferred scheme which 
is optimal from a physics point of view. 
A closely related question is how
to choose the renormalisation scale which is important since most
QCD observables are only known to next-to-leading order (NLO)
where the renormalisation scale dependence is still sizable.

Another problem with perturbative QCD predictions is that the series 
is in fact asymptotic, {\it i.e.} after a given order the higher order
contributions start to increase and make the series divergent. The most
prominent source for this asymptotic behaviour is due to so-called
renormalons which make the higher order coefficients grow 
factorially~\cite{renormalons}.

This talk presents an alternative approach which avoids, 
or at least minimizes, the problems outlined above by using 
conformal expansions and the closely related skeleton expansion. 
The presentation is mainly based on~\cite{BGGR} which also contains a
complete list of references. The relation between the skeleton expansion
and the Banks-Zaks expansion~\cite{Banks-Zaks}, as well as
the BLM scale-setting method by Brodsky, Lepage, and Mackenzie~\cite{BLM} 
and its generalizations~\cite{LM,CSR,BGKL,BLM-generalizations}, is also
discussed.

\section{Conformal relations}
For definiteness and simplicity the discussion will be limited to 
single-scale space-like observables in massless QCD, but the approach 
can also be generalised to time-like and multi-scale observables.
The perturbative expansion for such a single-scale 
observable can be written as,
$$
R(Q^2) = R_{\mbox{\tiny QPM}}(Q^2) + R_0(Q^2)\frac{\alpha_s(\mu^2)}{\pi}
           + R_1(Q^2,\mu^2)\frac{\alpha_s^2(\mu^2)}{\pi^2}
           + R_2(Q^2,\mu^2,\beta_2)\frac{\alpha_s^3(\mu^2)}{\pi^3}
           + \cdots \, ,
$$
where $Q^2=-q^2$ is the (space-like) physical scale,
$\mu^2$ is the renormalisation scale, and $\beta_2$ is the
next-to-next-to-leading order coefficient
in the renormalisation group equation for the coupling,
$$
\frac{d a(\mu^2)}{d \log(\mu^2)} = -\beta_0a^2(\mu^2)
           - \beta_1a^3(\mu^2)
           - \beta_2a^4(\mu^2)
           + \cdots \, .
$$
where $a=\alpha_s/\pi$.

The truncation of
the perturbative expansion at order $N$ introduces a renormalisation scale 
and scheme  uncertainty of order $a^{N+1}$. In addition 
the perturbative coefficients $R_n$ will asymptotically 
grow factorially due to renormalons, $R_n \sim n!\beta_0^n$.
This should be contrasted with the situation in  the conformal 
(scale-invariant) limit where $d a/d \log(\mu^2) = 0$. In this case
there is no scale-ambiguity, and the coefficients $R_n$ are
free of factorial growth due to renormalons. The only remaining problem
is the scheme uncertainty which can be circumvented by relating observables 
to each other instead of trying to make absolute predictions. 

Before continuing it is
useful to recall the concept of an effective charge~\cite{ECH} 
which collects all perturbative corrections to an observable. 
An observable $R(Q^2)$
can then be written in terms of the effective charge $a_R(Q^2)$ as,
$$
R(Q^2) = R_{\mbox{\tiny QPM}}(Q^2) + R_0(Q^2)a_R(Q^2) 
$$
where
$$ a_R(Q^2) = a(\mu^2)
           + r_1(Q^2,\mu^2) a^2(\mu^2) 
           + r_2(Q^2,\mu^2,\beta_2) a^3(\mu^2)
           + \cdots \, 
$$
and the perturbative coefficients $r_i = R_i/R_0$.

The most celebrated example of a conformal relation between observables
is the Crewther relation~\cite{Crewther,Crewther-2loop,BGKL} 
between the Adler D-function
($a_D$) and the polarized Bjorken sum-rule for deep inelastic scattering
($a_{g_1}$),
$$
(1+a_{D})(1-a_{g_1}) = 1 \, .
$$
Thus, the Crewther relation is simply a geometric series to all orders
and there is no growth of higher order coefficients.
The effective charges $a_{D}$ and $a_{g_1}$, which appear in the relation,
are defined by,
\begin{eqnarray*}
&& \hspace*{-1cm} D(Q^2)= Q^2\frac{d\Pi(Q^2)}{dQ^2} \equiv 
N_C \sum_f e_f^2\left[1+a_D(Q^2)\right]
\\ && \hspace*{-1cm}
\int_0^1 \left[g_1^p(x,Q^2)-g_1^n(x,Q^2)\right]dx\equiv
\frac{g_A}{6g_V}\left[1-a_{g_1}(Q^2)\right] 
\end{eqnarray*}
where $\Pi(Q^2)$ is the hadronic correction to the vacuum polarisation 
of the photon, the spacelike continuation of $R_{e^+e^-}(s)$.

In general, conformal relations between two arbitrary observables 
$A$ and $B$ can be written as,
$$
a_{A} = \sum_n c_n^{AB} a^n_{B}
$$
where, as is evident, the conformal coefficients 
$c_n^{AB}$ depend on which two observables that are related.
Of course, in real life the QCD coupling is scale-dependent.
Even so, the notion of conformal coefficients is still useful
as will be shown below. The main advantage is that by identifying 
the conformal part  of the ordinary perturbative coefficients 
 it is possible to treat all the running coupling effects 
separately and thus keeping the coefficients free from factorial 
growth due to renormalons which are instead resummed in the running of the
coupling.


\section{The skeleton expansion}

\begin{figure}[htb]
\begin{center}
\epsfig{file=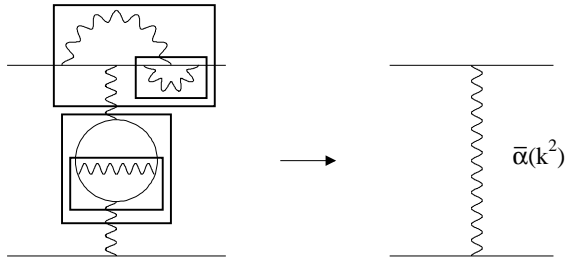,height=4cm}
\caption[0]{Example of an ordinary Feynman diagram (left) and the 
corresponding skeleton graph (right) in QED.}
\label{fig:qedskeleton}
\end{center}
\end{figure}

The skeleton expansion~\cite{Bjorken-Drell} organizes
the perturbative series in terms of contributions to fundamental
skeleton graphs. A skeleton graph is defined by the requirement 
that the fundamental 
vertices and propagators contain no substructure. One example of 
an ordinary Feynman diagram and the corresponding skeleton graph
is shown in Fig.~\ref{fig:qedskeleton}.

In QED, the skeleton expansion is straight-forward to construct 
thanks to the basic Ward identity, $Z_1=Z_2$, from which it 
follows that charge renormalisation is given by  photon
propagator renormalisation ($Z_3$). The coupling $\bar{\alpha}$
that appears in the skeleton expansion is the Gell-Mann Low coupling
which resums the Dyson series of the one-particle irreducible
photon self-energy $\Pi$,
$$
\bar{\alpha}(Q^2) = \frac{\alpha_0}{1-\Pi(Q^2)} \, .
$$
The radiative corrections to the one-photon exchange skeleton graph, 
such as illustrated
in Fig.~\ref{fig:qedskeleton}, can then be written as an integral
over the running coupling,
$$
\int \! \bar{a}(k^2)\, \phi_0\left(\frac{k^2}{Q^2}\right) 
\frac{dk^2}{k^2} 
$$
where $\phi_0$ is a  momentum distribution function which
has been normalised to 1 for convenience. 
(In the above example $\phi_0\left(\frac{k^2}{Q^2}\right) =\delta(k^2-Q^2)$.)

Adding the contributions from one-, two-, three-photon exchange etc.,
an effective charge can be written as
\begin{eqnarray}
a_R(Q^2)& = & 
\int \! \bar{a}(k^2) \,\phi_0\left(\frac{k^2}{Q^2}\right) 
\frac{dk^2}{k^2} +
\bar{c}_1\int \! \bar{a}(k_1^2) \bar{a}(k_2^2)
\,\phi_1\left(\frac{k_1^2}{Q^2},\frac{k_2^2}{Q^2}\right) 
\frac{dk_1^2}{k_1^2} \frac{dk_2^2}{k_2^2} \nonumber \\ & + &
\bar{c}_2\int \! \bar{a}(k_1^2) \bar{a}(k_2^2) \bar{a}(k_3^2)
\, \phi_2\left(\frac{k_1^2}{Q^2},\frac{k_2^2}{Q^2},\frac{k_3^2}{Q^2}\right) 
\frac{dk_1^2}{k_1^2} \frac{dk_2^2}{k_2^2} \frac{dk_3^2}{k_3^2} + \cdots
\label{eq:skeleton} \, ,
\end{eqnarray}
where $\phi_i$ are the
momentum distribution functions (normalised to 1)
and the $\bar{c}_i$ are the conformal coefficients in the skeleton scheme.
For simplicity the above expression has been written including just one
skeleton at each order but in general there can be several 
different skeletons which contribute at the same order. For comparison,
the conformal theory gives
$a_R(Q^2) = \bar{a} + \bar{c}_1 \bar{a}^2 + \bar{c}_2 \bar{a}^3  + \cdots $.

Another important property of the skeleton expansion is that each term
 in the expansion is renormalisation scheme and scale-invariant 
by itself. In addition the skeleton coupling is gauge-invariant.
The skeleton expansion thus provides an alternative way of writing the 
perturbative series for an observable in which each term is given by
one or several integrals over the running coupling. One complication
of the skeleton expansion is that in general one needs a diagrammatic 
construction to identify the different skeletons. However, at low orders 
this requirement can be bypassed.

\begin{figure}[htb]
\begin{center}
\epsfig{file=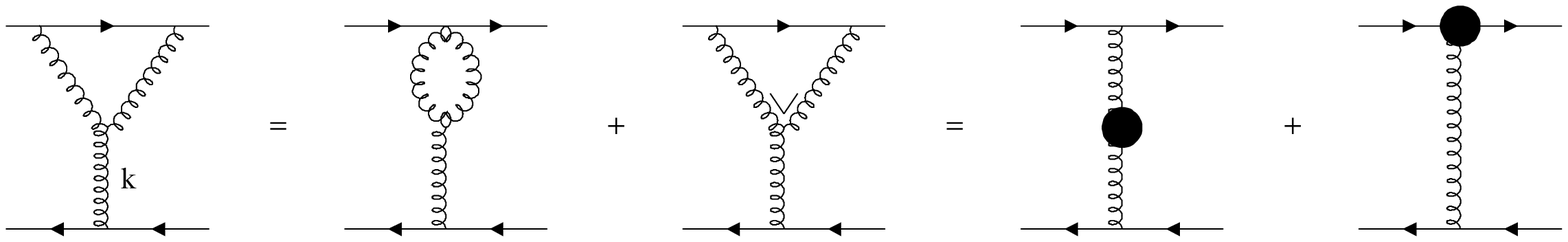,height=2.5cm}
\caption[0]{Illustration of the subdivision of the
three-gluon vertex into a pinch part and a non-pinch part 
using the pinch technique}
\label{fig:3gluon}
\end{center}
\end{figure}

In QCD, the existence of an all-order skeleton expansion has so far not been
proved. The basic complication arises from the gluon self-interactions
and the related difference between gluon-propagator and charge renormalisation.
Nevertheless it is reasonable to assume that something similar to 
the skeleton expansion in QED can also be constructed for QCD. In fact,
the so called pinch technique~\cite{Pinch} provides a realisation 
of the skeleton expansion in QCD at the one-loop level. As an example
Fig.~\ref{fig:3gluon} illustrates how the three-gluon vertex is divided
into a pinch part which contributes to the renormalisation of the 
effective propagator and a non-pinch part which contributes to 
renormalisation of the ``external'' vertex.

In this way the pinch technique arrives at the following QED-like Ward 
identities,
\begin{eqnarray*}
Z_1^{(PT)} &=&  Z_2^{(PT)} = 1-\frac{1}{\varepsilon}\frac{C_F}{4}\bar{a} \\
Z_3^{(PT)} &=& 1+\frac{1}{\varepsilon}
\left(\frac{11}{12}C_A-\frac{1}{3}T_FN_F\right) \bar{a} =
1+\frac{1}{\varepsilon}\beta_0 \bar{a} 
\end{eqnarray*}
such that all the one-loop running coupling effects are contained
in the effective gluon propagator.
The coupling defined by the pinch technique has a simple relation to
the $\overline{\mbox{MS}}$ scheme,
$$
\bar{a}(Q^2) = a_{\overline{\mbox{\tiny MS}}}(\mu^2) 
+ \left[ -\beta_0\left(\log\frac{Q^2}{\mu^2}-\frac{5}{3}\right)+1 \right]
a^2_{\overline{\mbox{\tiny MS}}}(\mu^2) +\cdots \, .
$$
Recently there has been progress in extending the pinch-technique
to two loops~\cite{Pinch2} and this may eventually lead to an extension
of the skeleton expansion in QCD to two loops as well.
Another possibility may be to use light-front quantization of QCD
in light-cone gauge~\cite{Prem}.


\section{Identifying conformal coefficients}

Given the advantages of the skeleton expansion compared to the standard
perturbative expansion, it is instructive to consider the following 
simplified ansatz for QCD as a starting point for further investigations:
assume there is only one skeleton coupling, that
 there is only one skeleton graph at each order in $\bar{a}$, and that
the dependence on the number of light flavours ($N_F$) can be used
to identify the non-conformal parts of the perturbative coefficients.
Given these assumptions the first conformal coefficients in the skeleton
expansion can be obtained from the perturbative ones in the following
way~\cite{BGGR}.

The starting point is the skeleton expansion of an effective charge
given by Eq.~(\ref{eq:skeleton}).
Next the skeleton couplings $\bar{a}(k^2)$ 
under the integration sign can be expanded
in the coupling $\bar{a}(Q^2)$ using the solution to 
the renormalisation group equation,
$$
\bar{a}(k^2) = \bar{a}(Q^2)
+\beta_0\log\left(\frac{Q^2}{k^2}\right)\bar{a}^2(Q^2) 
+\left[\beta_1\log\left(\frac{Q^2}{k^2}\right)
       +\beta_0^2\log^2\left(\frac{Q^2}{k^2}\right)\right]\bar{a}^3(Q^2)
+\cdots \, .
$$
Inserting this into Eq.~(\ref{eq:skeleton}) then gives,
$$
a_R(Q^2) =
 \bar{a}(Q^2)
 + \left(\bar{c}_1+\beta_0\phi_0^{(1)}\right) \bar{a}^2(Q^2) 
 + \left(\bar{c}_2+\bar{c}_1\beta_0\phi_1^{(1)}
         +\beta_1\phi_0^{(1)}+\beta_0^2\phi_0^{(2)}\right)\bar{a}^3(Q^2) 
 + \cdots 
$$
where $\phi_i^{(n)}$ are log-moments of the momentum distribution functions,
\begin{eqnarray*}
  \phi_0^{(n)} & = &
\int \! \log^n\left(\frac{Q^2}{k^2}\right) \phi_0\left(\frac{k^2}{Q^2}\right) 
\frac{dk^2}{k^2} \\
\phi_1^{(1)} & = & 
\int \! \left( \log\frac{Q^2}{k_1^2} + \log\frac{Q^2}{k_2^2}\right) 
\phi_1\left(\frac{k_1^2}{Q^2},\frac{k_2^2}{Q^2}\right) 
\frac{dk_1^2}{k_1^2} \frac{dk_2^2}{k_2^2} \, .
\end{eqnarray*}
This can now be directly compared with the standard perturbative expansion,
$$
a_R(Q^2) =
 \bar{a}(Q^2) + \bar{r}_1\bar{a}^2(Q^2)  +\bar{r}_2\bar{a}^3(Q^2) + \cdots \, ,
$$
which gives the relations
\begin{eqnarray*}
 \bar{r}_1  & = &\bar{c}_1+\beta_0\phi_0^{(1)}
  \\
\bar{r}_2  & = & \bar{c}_2+\bar{c}_1\beta_0\phi_1^{(1)}
         +\beta_1\phi_0^{(1)}+\beta_0^2\phi_0^{(2)}     \, .
\end{eqnarray*}
Based on the $N_F$ dependence of the perturbative coefficients  $\bar{r}_i$
it is thus possible to identify in a unique way the conformal coefficients 
$\bar{c}_1$ and $\bar{c}_2$ as well as the log-moments $\phi_0^{(1)}$, 
$\phi_0^{(2)}$, and $\phi_1^{(1)}$. (This follows since the coefficients 
$\bar{r}_i$ are polynomials in $N_F$ of order $i$.)
In fact, given the assumptions made,
it is possible to decompose the perturbative coefficients up to order 
$\bar{a}^4$ without any additional information. At higher orders the $N_F$
dependence alone does not provide enough information even with the
simplifying assumptions that have been made.

In general there are several ways in which the assumed ansatz can
break down. Most notably, at higher orders there are skeletons which are 
$N_F$-dependent by themselves. In contrast to QED where the $N_F$-dependent 
skeletons (such as the light-by-light scattering diagrams) can be easily
identified based on the dependence on the external charge there is in general
no such simple identification possible in QCD. Another complication is that
there may be more than one skeleton at each order. To resolve these two
problems one will need an explicit diagrammatic construction of the 
skeleton expansion. It may also be the case that the skeleton expansion in QCD
can only be systematically extended to all orders by having several skeleton
couplings. However, even if some of the assumptions that have been made are
wrong, it may still be true that the general properties of the
ansatz are valid. This includes the property that running-coupling effects
can be associated with different skeleton graphs in a 
renormalisation-group-invariant way, 
and that the skeleton coefficients are conformal. In practice
there is usually no problem in identifying the skeleton structure at 
next-to-leading order but special care has to be taken 
as will be discussed below when the application of BLM scale-setting to the
thrust-distribution in $e^+e^-$-annihilation is re-examined.

\section{Relation to the Banks-Zaks expansion}
As already realised at the time of the discovery of asymptotic freedom, 
perturbative QCD has an perturbative infrared 
fixed-point~\cite{infrared} 
($k^2 \to 0$),
$$
\frac{d a_{\mbox{\tiny FP}}(k^2)}{d\ln k^2}   = 
-\beta_0 a^2_{\mbox{\tiny FP}}(k^2) 
 -\beta_1 a^3_{\mbox{\tiny FP}}(k^2)+\cdots = 0
$$
in the so called conformal window $8<N_F<16$ since for this 
range of $N_F$ the first two terms in the $\beta$-function have opposite 
signs, $\beta_0 = \frac{11}{4}-\frac{1}{6}N_F>0$ 
and  $\beta_1 = \frac{51}{8}-\frac{19}{24}N_F<0$.

If the coupling at the fixed point $a_{\mbox{\tiny FP}}$ is small,
 such that perturbation
theory is still applicable, then it can be written as a so called 
Banks-Zaks expansion~\cite{Banks-Zaks} in the parameter
$a_0=-\beta_0/\left.\beta_1\right|_{\beta_0=0}=\frac{16}{107}\beta_0$, 
$$
 a_{\mbox{\tiny FP}} = a_0+v_1a_0^2+\cdots   \, ,
$$
where the coefficients $v_i$ can be calculated from the higher order terms 
($\beta_2$ etc.) in the $\beta$-function.

In the same way an arbitrary effective charge $a_R$ can also be expanded in
$a_0$. Starting from the ordinary perturbative expansion the coefficients
$r_i$ can be rewritten in terms of $a_0$ using the polynomial 
$N_F$-dependence,
$$
a_R(Q^2) = a(Q^2) +(r_{1,0}+r_{1,1}a_0)a^2(Q^2) 
+(r_{2,0}+r_{2,1}a_0+r_{2,2}a_0^2)a^3(Q^2) + \cdots \, .
$$
From this it follows that it is also possible to get a relation between
the fixed-point value of the effective charge $a_R^{\mbox{\tiny FP}}$
and the coupling $a_{\mbox{\tiny FP}}$. 
Taking the limit $Q^2 \to 0$ (assuming that this is well defined)
and inserting $a_0=a_{\mbox{\tiny FP}} +u_1a_{\mbox{\tiny FP}}^2 +\cdots$ 
gives the fixed point relation,
\begin{eqnarray*}
a_R^{\mbox{\tiny FP}} & = &  
a_{\mbox{\tiny FP}} +r_{1,0}a_{\mbox{\tiny FP}}^2
+(r_{2,0}+r_{1,1})a_{\mbox{\tiny FP}}^3 + \cdots \, .
\end{eqnarray*}
Comparison with the conformal coefficients obtained from the 
skeleton decomposition of the perturbative coefficients shows that,
if $a_{\mbox{\tiny FP}}$ is identified with the skeleton coupling then,
they are indeed the same, {\it i.e.} $r_{1,0}=\bar{c}_1$
and $r_{2,0}+r_{1,1}=\bar{c}_2$ etc. Thus, the conformal
coefficients in QCD can also obtained from the Banks-Zaks expansion
by analytically continuing the number of light quark flavours into the
conformal window and taking the infrared limit~\cite{BGGR}.





\section{Connection to BLM scale-setting}

Once the conformal coefficients have been identified one also has to evaluate
the corresponding skeleton integrals. For the leading skeleton this can be done
using the momentum distribution function calculated in the large
$\beta_0$-approximation. At the same time the associated renormalon ambiguity
indicates the form of the non-perturbative corrections in terms of
power-corrections. The combination gives a framework for analysing the
renormalon resummation and the non-perturbative corrections
together~\cite{Grunberg,GG}.
  An alternative is to approximate the skeleton integrals by
using BLM scale-setting~\cite{BLM,LM} as will be discussed below.

The starting point is the skeleton expansion of the effective charge
in question where each integral is evaluated using the mean value theorem (MVT)
in the following way,
\begin{eqnarray*}
  a_R(Q^2) & = &
\int \! \bar{a}(\ell^2) \, \phi_0\left(\frac{\ell^2}{Q^2}\right) 
\frac{d\ell^2}{\ell^2} 
+
\bar{c}_1\int \! \bar{a}(\ell_1^2) \bar{a}(\ell_2^2)
 \, \phi_1\left(\frac{\ell_1^2}{Q^2},\frac{\ell_2^2}{Q^2}\right) 
\frac{d\ell_1^2}{\ell_1^2} \frac{d\ell_2^2}{\ell_2^2}
 \\ & + &
\bar{c}_2 \int \! 
\bar{a}(\ell_1^2) \bar{a}(\ell_2^2) \bar{a}(\ell_3^2)
 \, \phi_2\left(\frac{\ell_1^2}{Q^2},\frac{\ell_2^2}{Q^2},\frac{\ell_3^2}{Q^2}\right) 
\frac{d\ell_1^2}{\ell_1^2} \frac{d\ell_2^2}{\ell_2^2} \frac{d\ell_3^2}{\ell_3^2} +\cdots 
 \\ (\mbox{\small MVT}) & \equiv & 
\bar{a}(k_0^2) + \bar{c}_1 \bar{a}^2(k_1^2) + \bar{c}_2 \bar{a}^3(k_2^2)
+\cdots
\end{eqnarray*}
The ``BLM'' scales $k_0$, $k_1$, $k_2$, etc. are uniquely 
determined by requiring a one-to-one correspondence between 
the skeleton integrals and the terms in the ``BLM'' series~\cite{BGGR}. 
In other words $k_0$ depends only on $\phi_0$, $k_1$ on $\phi_1$, 
and so on. Thus there is no ambiguity in determining the
scales as is the case for commensurate scale relations~\cite{CSR,BGKL}.
Expanding the couplings $\bar{a}(k^2)$ in terms of $\bar{a}(Q^2)$ 
under the integration sign the ``BLM'' 
scales are obtained as a perturbative series in the skeleton 
coupling with the coefficients given in terms of the moments
of the distribution functions,
\begin{eqnarray*}
\ln\frac{Q^2}{k_0^2} & = & 
\phi_0^{(1)} + 
\left[\phi_0^{(2)}-\left(\phi_0^{(1)}\right)^2\right]\beta_0\bar{a}(k_0^2)
+\cdots \, ,
\\&& \mbox{\tiny mean}  \hspace*{1.3cm}\mbox{\tiny variance} 
\\
\ln\frac{Q^2}{k_1^2} &=& \frac{1}{2}\phi_1^{(1)} +\cdots  \, .
\end{eqnarray*}
It is important to realize that this 
provides a systematic improvement of the original BLM-scale,
$k_{0,\mbox{\tiny BLM}}^2=Q^2\exp\left(-\phi_0^{(1)}\right)$. In the lowest
order approximation the scale $k_0$ is simply given by the mean of the 
momentum distribution as indicated above. By going to higher orders one then
takes into account the variance of the distribution and so on. This corresponds
to performing the skeleton integral with successively improved approximations
to $\phi_0$.

Given the conformal expansions of two observables in the skeleton scheme
it is also possible to eliminate the skeleton scheme and get a direct relation
between the two observables -- a so called commensurate scale relation (CSR).
From renormalisation group transitivity it follows that the coefficients
in the commensurate scale relation are also conformal and thus free
of factorial growth due to renormalons. However, there
is no clear interpretation of the scales that appear in the CSRs,
and in addition there is no unique scale setting procedure as has been
already mentioned.


\section{Re-examining BLM scale-setting for thrust}

The new insights gained from the relation between the skeleton expansion
and BLM scale-setting makes it interesting to re-examine BLM scale-setting
for event shape observables in $e^+e^-$ annihilation~\cite{eeblm}. 
In the following the
thrust distribution will be considered as a concrete example but 
general criteria for the applicability of BLM scale-setting will also
be given.

\begin{figure}[ht]
\begin{center}
\epsfig{file=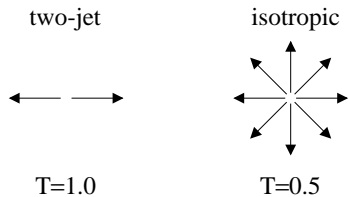,height=3cm}
\caption[0]{The values of thrust for an event with two 
narrow back-to-back jets (left)
and an event with isotropic distribution of particles (right).}
\label{fig:thrust}
\end{center}
\end{figure}

Thrust is an event shape observable defined by,
$$
T = \max_{\vec{n}_T} 
\frac{\sum_i \vec{n}_T \cdot \vec{p}_i}
     {\sum_i |\vec{p}_i|}
$$
where the sum runs over all particles in the final state.
The thrust-axis $\vec{n}_T$ is varied until 
the maximal value for $T$ is obtained. 
An event with two narrow back-to-back jets corresponds to $T=1$
whereas the minimal thrust value $T=0.5$ is obtained for
an event with isotropic distribution of particles 
as illustrated in Fig.~\ref{fig:thrust}.

In the quark parton model the thrust distribution is  a 
delta-function at $T=1$. The leading order QCD-corrections have been
calculated analytically~\cite{thrust-lo} whereas the next-to-leading order 
QCD-corrections have only been calculated numerically~\cite{thrust-nlo,beowulf}. 
From the definition of thrust one expects that at leading order there 
is only one skeleton which contributes and that all the $N_F$-dependence 
at next-to-leading order is from running coupling effects. 
Thus BLM scale-setting should be straight forward.
There is however one possible complication, namely the non-inclusiveness
of the definition. The easiest way to see this is that at leading order
thrust can have values in the range $2/3 < T<1$ whereas at next-to-leading 
order the range is given by $1/\sqrt{3} < T<1$. Thus,
if the next-to-leading order $N_F$-dependence is non-zero for 
$1/\sqrt{3} < T< 2/3$ then this cannot be attributed to the leading skeleton.
However, as will be shown below the problems for the case of thrust 
are minimal.

At next-to-leading order the BLM series for the thrust-distribution
can be written as,
\begin{equation}
\label{eq:conf}
\frac{1}{\sigma} \frac{d\sigma^{\mbox{\small BLM}}}{dT}(s,T) =
\delta(1-T)
 + R_0(T) \,  \bar{a} \! \left(k_{0,\mbox{\tiny BLM}}^2(s,T)\right)
 + \bar{R}_1(T)  \, \bar{a}^2 \! \left(k_{0,\mbox{\tiny BLM}}^2(s,T)\right) \, ,
\end{equation}
where the skeleton coupling has been identified with the pinch technique
coupling, $\bar{R}_1(T)$ is the conformal coefficient in the pinch 
scheme, and the BLM-scale $k_{0,\mbox{\tiny BLM}}^2$ is used
to approximate the unknown scale $k_1^2$ which should appear in  
the $\bar{R}_1(T)$-term.
It is important to realize that the BLM-scale $k_{0,\mbox{\tiny BLM}}^2(s,T)$ 
is a function of both kinematic variables, $s$ and $T$. In addition
the BLM scale is undefined for $T<2/3$ where $R_0$ vanishes.

\begin{figure}[t]
\epsfig{file=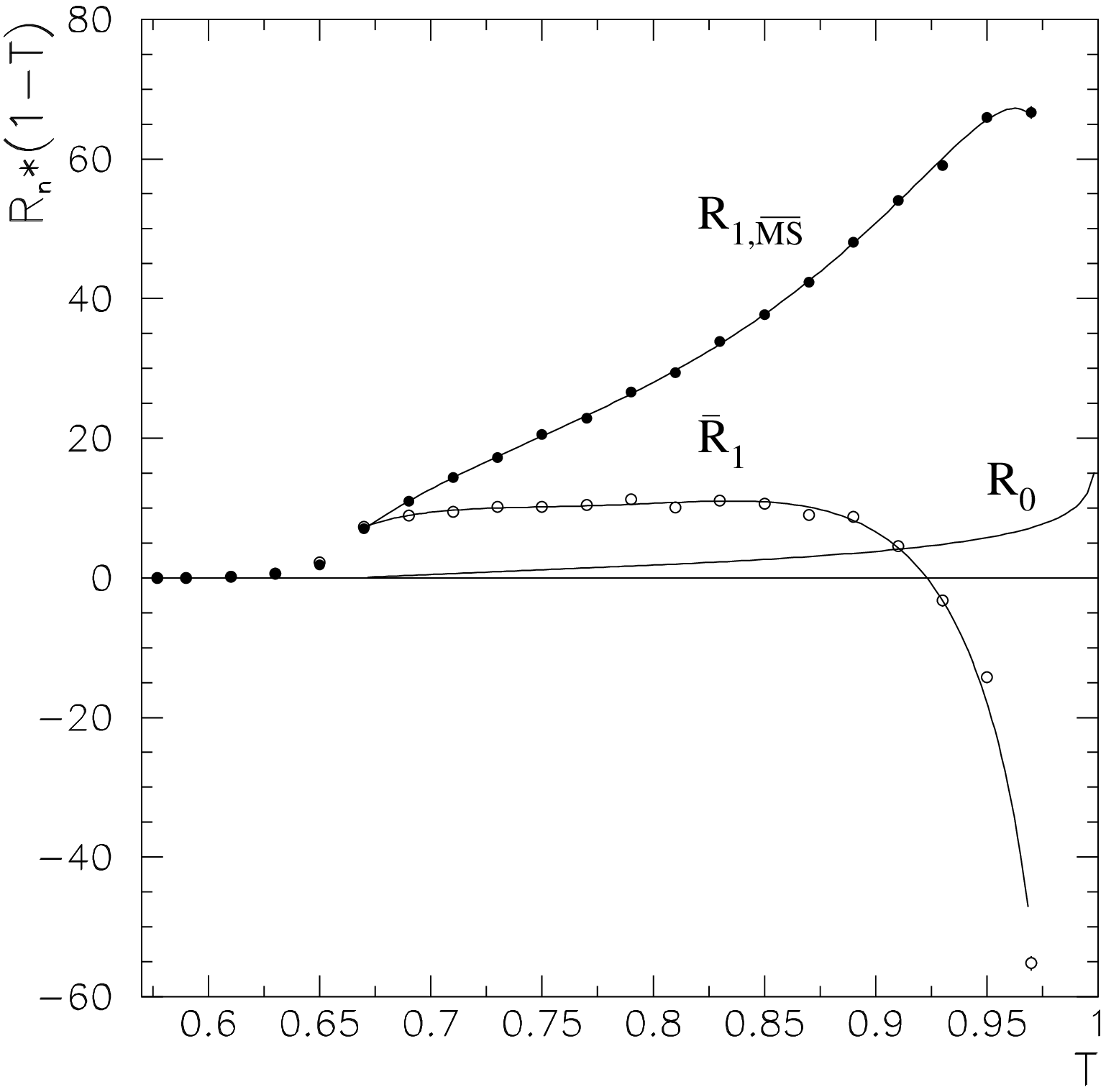,height=7.5cm}
\epsfig{file=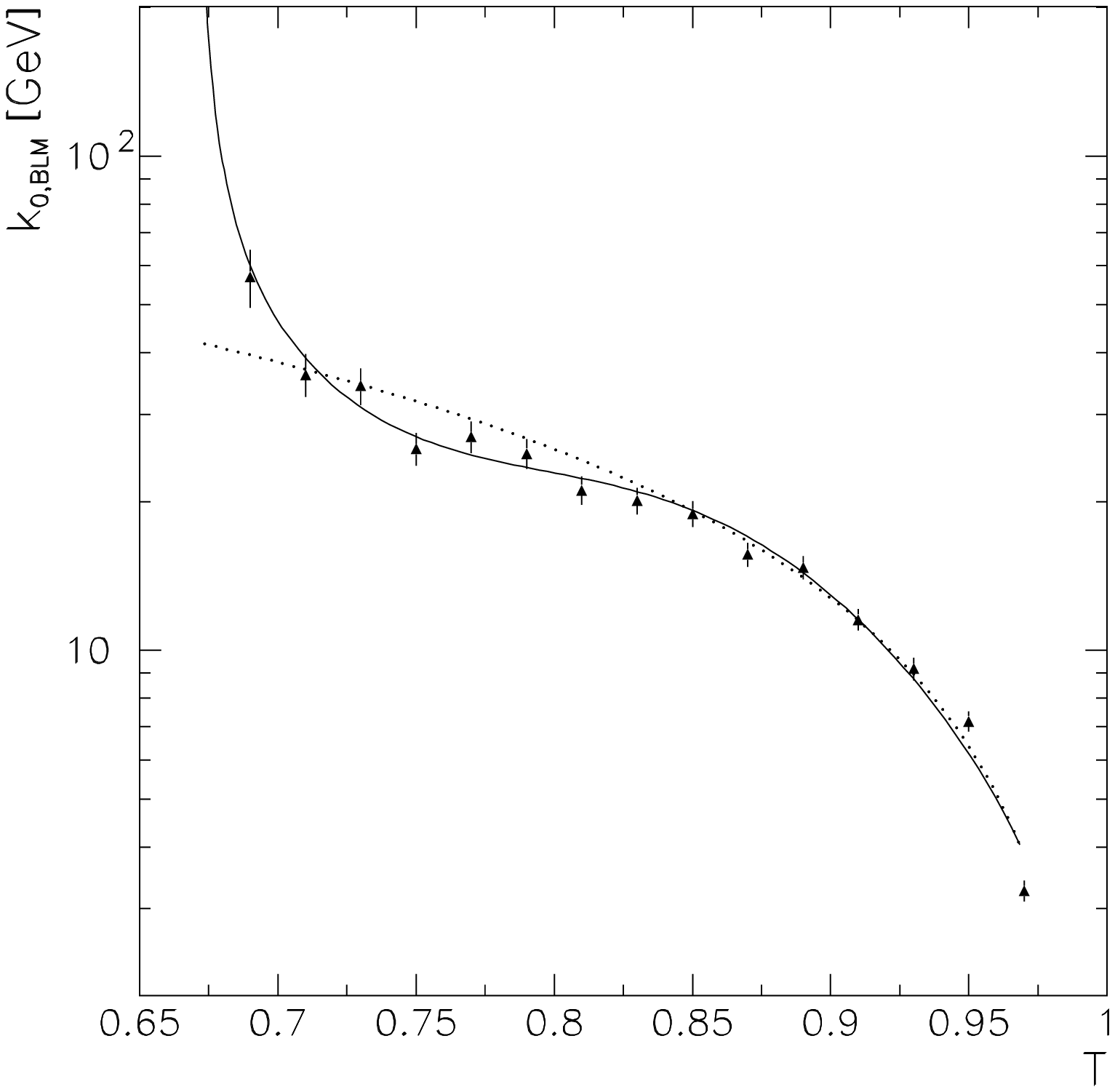,height=7.5cm}
\caption[0]{(a) The next-to-leading order 
conformal coefficient $\bar{R}_1$ compared to the standard 
$\overline{\mbox{MS}}$ coefficient 
$R_{1,\overline{\mbox{\tiny MS}}}(\mu^2=s,N_F=5)$
and the leading order (scheme-invariant) coefficient $R_0$. 
(b) The BLM-scale for $\sqrt{s}=M_Z$.
For both figures the points show the numerical values that have been
calculated and the lines are fits to these points taking into account the 
known logarithmic terms. In (b) the dotted line show the approximation
$k_{0,\mbox{\tiny BLM}} \simeq 1.4(1-T)\sqrt{s}$.}
\label{fig:coeff}
\end{figure}

The expansion given above should be compared with the standard 
$\overline{\mbox{MS}}$ expansion using $\mu^2=s$,
$$
\frac{1}{\sigma} \frac{d\sigma^{\overline{\mbox{\tiny MS}}}}{dT}(s,T) =  
\delta(1-T)
 + R_0(T) a_{\overline{\mbox{\tiny MS}}}(s) 
 + R_{1,\overline{\mbox{\tiny MS}}}(\mu^2=s,N_F,T) 
 a^2_{\overline{\mbox{\tiny MS}}}(s) \, .
$$
The leading order coefficient $R_0$ is scheme-invariant and thus the same in
both expansions. However, the next-to-leading order coefficient $R_1$ is very
different in the two cases as is illustrated in Fig.~\ref{fig:coeff} which
shows the conformal coefficient $\bar{R}_1$ compared to the standard 
$\overline{\mbox{MS}}$ coefficient 
$R_{1,\overline{\mbox{\tiny MS}}}(\mu^2=s,N_F=5)$
and the leading order coefficient $R_0$. The coefficients
have been calculated numerically using the Beowulf program~\cite{beowulf} 
which is shown as points in the figures. The lines are fits to this points
taking into account the know logarithmic parts of the
coefficients~\cite{nll}.

From the figure it is clear that the next-to-leading order coefficient is large
compared to the leading order one in both cases. However, the conformal
coefficient is more stable over a large range of $T$ (when multiplied with
$(1-T)$) except for $T \to 1$ where it becomes negative. 
This is the Sudakov region which can 
only be properly treated by resumming all singular terms in the
Sudakov form-factor. Another important feature which is clear from the figure
is that the non-conformal part of $R_1$ more or less vanishes for $T<2/3$,
which is a good indication that the $N_F$ dependence can indeed be used to 
separate the conformal and non-conformal parts and
that the problems with non-inclusiveness are only minor 
(see also~\cite{NS,GG,EG}). This property is
different for other event shape observables depending on how they are
defined. For example, oblateness is defined as the difference between an
observable that starts at order $\alpha_s$ and one that starts at order 
$\alpha_s^2$. As a consequence there are $N_F$ dependent
contributions to the next-to-leading term which do not come from the leading
skeleton. This could also explain why BLM scale-setting seems to fail for
some event shape observables~\cite{eeblm}.

\begin{figure}[t]
\epsfig{file=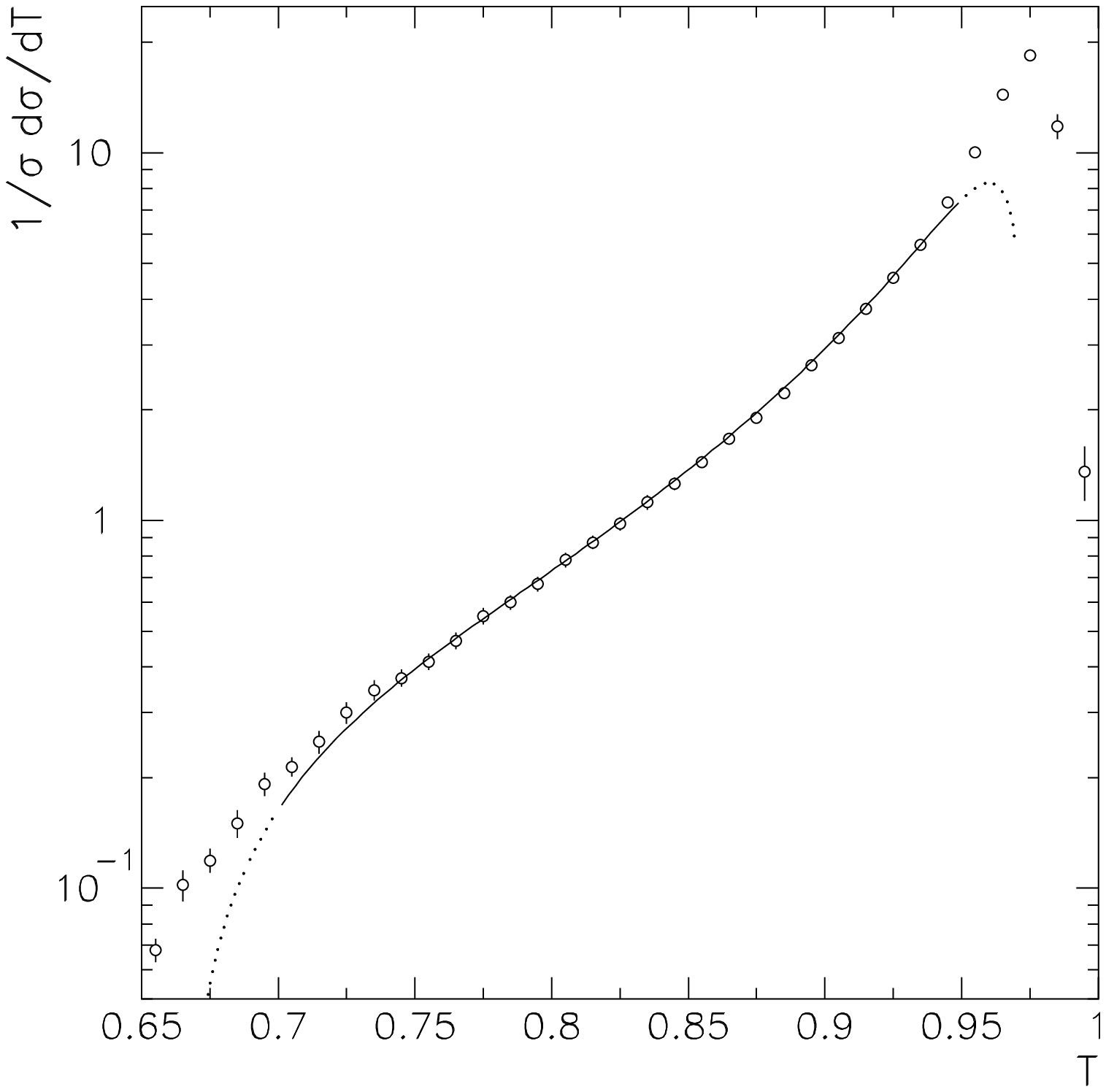,height=7.5cm}
\epsfig{file=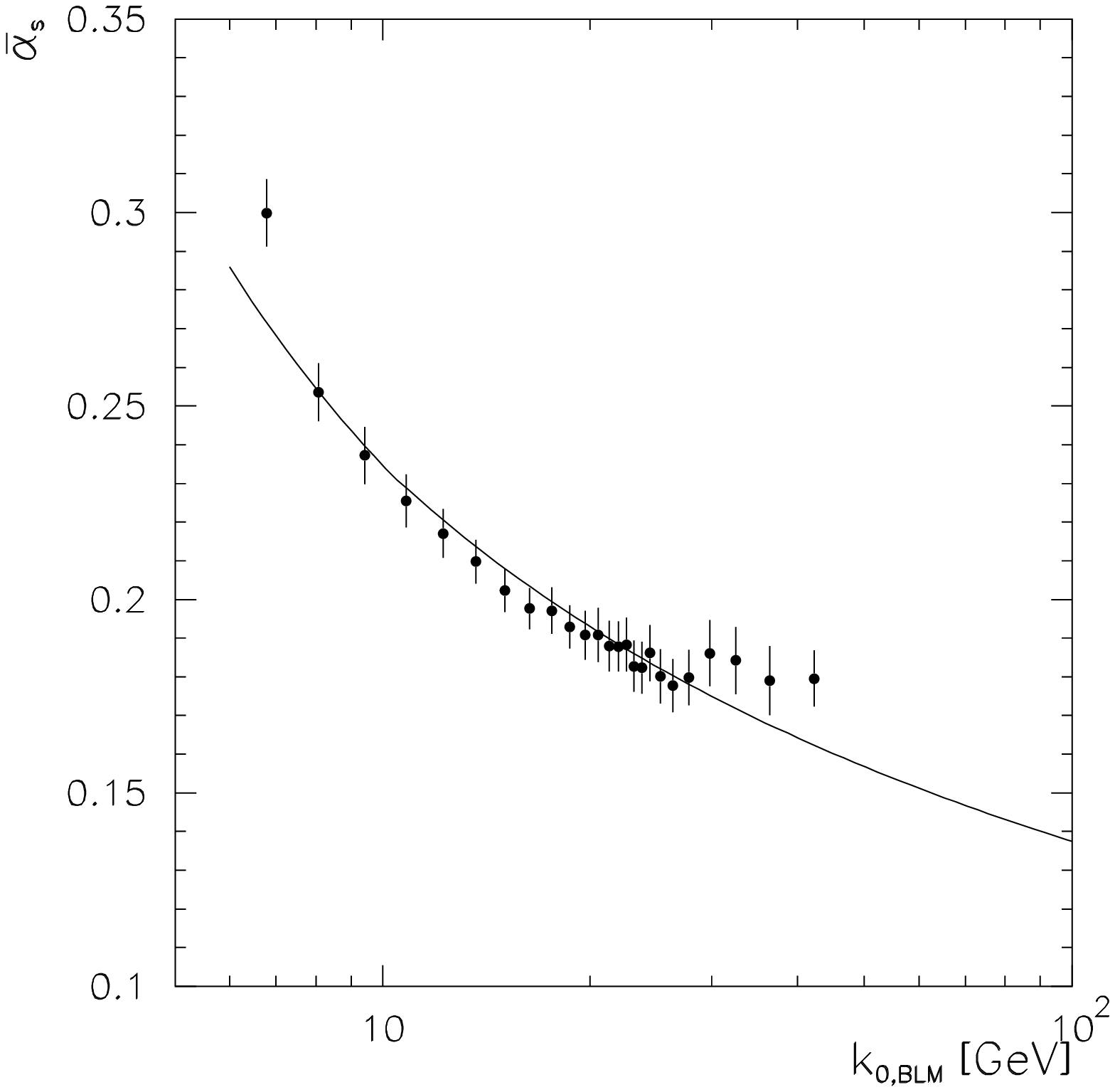,height=7.5cm}
\caption[0]{(a) Fit to OPAL data using the fixed order BLM series. The
full line corresponds to the range fitted ($0.7<T<0.95$).
(b) Value of running coupling $\bar{\alpha}_s$ extracted from OPAL 
data in the range $0.7<T<0.95$ at the corresponding BLM-scale.}
\label{fig:opal}
\end{figure}

Fig.~\ref{fig:coeff} also shows the resulting BLM-scale 
$k_{0,\mbox{\tiny BLM}}(s,T)$ for the case $\sqrt{s}=M_Z$.
From the figure it is clear that the scale vanishes as $T \to 1$
which is reasonable since the available phase-space for gluon emission
vanishes in this limit. For comparison the figure also shows the 
approximation $k_{0,\mbox{\tiny BLM}} \simeq 1.4(1-T)\sqrt{s}$ 
which gives an overall good description of the $T$-dependence.
The scale can also be understood physically as the 
transverse momentum which approximately scales as $(1-T)\sqrt{s}$
for a three-jet configuration with one of the jets being much less energetic
than the other two, {\it i.e.} in the soft limit.
For $T\to 2/3$ the BLM-scale grows rapidly since the $R_0 \to 0$
but even at $T=0.69$ (the point with the smallest $T$-value shown in 
the figure) the BLM-scale is still smaller than $\sqrt{s}$ which
should be true in a physical scheme following from the mean value theorem.

For illustration, the fixed order BLM expression for the thrust distribution
given by Eq.~(\ref{eq:conf}) has been fitted to data from the OPAL 
collaboration~\cite{opal} at $\sqrt{s}=M_Z$ in the range $0.70<T<0.95$ 
using a two-loop running coupling. The result of the fit, which is shown 
together with the data in Fig.~\ref{fig:opal}, 
corresponds
to the value $\alpha_{\overline{\mbox{\tiny MS}}}(M_Z^2)=0.117$.
(To translate the fit into a value for
$\alpha_{\overline{\mbox{\tiny MS}}}(M_Z^2)$ the
commensurate scale relation,
$\bar{a}(e^{5/3}M_Z^2) = a_{\overline{\mbox{\tiny MS}}}(M_Z^2) 
+ a^2_{\overline{\mbox{\tiny MS}}}(M_Z^2) $,  was used.) For comparison,
using the fixed order $\overline{\mbox{MS}}$ expression gives  
$\alpha_{\overline{\mbox{\tiny MS}}}(M_Z^2)=0.143$. This illustrates the 
importance of taking running coupling effects into account.
However, it should be kept in mind that a complete analysis should 
also include the Sudakov form-factor and non-perturbative effects.

It is also possible to see the running of the coupling $\bar{\alpha}_s$ 
as a function of the BLM-scale $k_{0,\mbox{\tiny BLM}}$ directly from
the data. For each data point Eq.~(\ref{eq:conf}) is a simple second order
equation which can be  solved for $\bar{a}=\bar{\alpha}_s/\pi$. The resulting
values of $\bar{\alpha}_s$ obtained in this way are shown in
Fig.~\ref{fig:opal} as a function of the corresponding BLM-scales.
(The figure only shows the points that where used in the fit. For larger values
of $T$ the next-to-leading order coefficient $\bar{R}_1$ is negative and
for smaller values of $T$ the next-to-leading order correction is larger than
100\%.) Thus, even though the experiment is done at a fixed energy, it is
still possible to observe the running of the coupling.


\section{Conclusions}
The standard perturbative expansion of observables in QCD is plagued by 
renormalisation scheme and scale ambiguities as well as higher order
coefficients which grow factorially due to renormalons. In this talk I have
presented an alternative approach which avoids,  or at least 
minimizes, these problems by using conformal expansions, 
especially the skeleton expansion.

In contrast to the ordinary perturbative expansion the skeleton expansion 
is free of renormalisation scheme and scale ambiguities and
the coefficients are free of factorial growth due to renormalons. 
Presently the pinch technique provides a realization of the skeleton expansion
in QCD at next-to-leading order but it is not known whether an all-order
expansion exists or not. Even so, the skeleton expansion has important
phenomenological consequences.

The leading skeleton integral makes it 
possible to include non-perturbative effects in a consistent
way which takes into account the arbitrariness of the
definition of perturbation theory. The renormalon ambiguities which
appear in the evaluation of the leading skeleton integral can be used to
parametrize the non-perturbative contributions in the form of
power-corrections. 

By making a simple ansatz for the skeleton expansion in QCD 
the first steps in making a more systematic study
of it properties have been taken~\cite{BGGR}. One result of this
study is that the conformal coefficients coincide with the
ones obtained in case QCD has a perturbative infrared fixed-point 
(the Banks-Zaks expansion).

The skeleton integrals which appear in the skeleton expansion  
can also be approximated by the BLM-scale setting method and 
its generalisations. Requiring a one-to-one
correspondence between the BLM-scales and the skeleton integrals gives
a unique prescription for setting the scales~\cite{BGGR} in contrast to the
situation for commensurate scale relations. 
The connection between the skeleton expansion and BLM scale-setting 
also gives new criteria for the applicability of the latter.

\Acknowledgments
I would like to thank Stan Brodsky and Einan Gardi for helpful
comments on the manuscript. I would also like to thank the organizers
for financial support and for organizing such an inspiring meeting.

\end{document}